\documentstyle[11pt,hyfestconf,epsf,twoside,graphics]{article}

\def \etal {{\em et al.}}
\def \eg {{\em e.g.}}
\def \ie {{\em i.e.}}
\def \cf {{\em c.f.}}

\begin{document}

\title{Near-Infrared Emission Line Searches for High-Redshift Galaxies}
\author{Andrew Bunker\altaffilmark{1}}
\affil{Dept.\ of Astronomy, University of California, Berkeley
CA~94720\\
{\tt email: bunker@ast.cam.ac.uk}}
\altaffiltext{1}{Present address: Institute of Astronomy, Madingley Road,
Cambridge, CB3~OHA, England}

\begin{abstract}
In this article I review recent developments in near-infrared emission
line searches for star-forming galaxies at high redshift.  Using the
$J$-, $H$- \& $K$-bands we can potentially chart the history of star
formation over the range $1<z<5$ using the prominent rest-optical
nebular emission lines alone, filling in the ``redshift desert'' at
$z\sim1-3$ where most common emission lines lie outside the optical
bands. Studying the rest-frame optical at $z\sim 2$ also allows a fair
comparison to be made with the local Universe -- the rest-optical lines
are vastly more robust to extinction by dust than the rest-UV, with the
resonantly-scattered Ly$\alpha$ line particularly unreliable. I discuss
the recent history of near-infrared emission-line searches using
narrow-band imaging and spectroscopy, and look to the future in this era
of 10\,m telescopes: such work gives us the potential to push to $z>10$,
the next frontier in the Hy-Redshift Universe.
\end{abstract}


\section{Introduction}

This conference celebrates some of the research interests of Hyron
Spinrad on his 65th birthday. Hy's career has been synonymous with the
discovery of the most distant galaxies, and in recent years this field
has undergone a renaissance. Galaxies at $>$90\% of the look-back time
are now being successfully hunted with a broad armory of observational
techniques. In this article I will review recent developments in
near-infrared emission-line searches.

The observational study of galaxy evolution aims to follow the star
formation, gas consumption, metal enrichment and merging rates of the
components which have combined to form today's galaxies.  In the last
four years, our knowledge of high redshift `normal' (non-AGN) galaxies
has blossomed. The photometric selection technique of Steidel, Pettini
\& Hamilton (1995), using the intrinsic Lyman limit continuum break at
$\lambda_{\rm rest}=91.2$\,nm and the blanketing effect of intervening
Lyman-$\alpha$ absorbers below $\lambda_{\rm rest}=121.6$\,nm, has
proven a robust way to select star-forming galaxies through broad-band
optical imaging at $z\approx 3$ (the `$U$-drops', Steidel \etal\
1996a,b) and now at $z\approx 4$ (the `$B$-drops', Steidel \etal\ 1999
and Chuck Steidel's contribution to this conference proceedings). In
addition, after many years of largely fruitless searches for
Ly$\alpha$\,121.6\,nm (\eg, Pritchet \& Hartwick 1990; de~Propris \etal\
1993; Thompson, Djorgovski \& Trauger 1995), a number of field galaxies
have now been identified at $z\sim 3-6$ through this emission line (\eg,
Dey \etal\ 1998; Hu, Cowie \& McMahon 1998; Chen \etal\ 1999; Steidel
\etal\ 2000 -- see also the articles by Ken Lanzetta, George Djorgovski,
Esther Hu and Hy Spinrad in this volume).

Observations in the optical are now, at last, revealing large numbers of
high-redshift galaxies. Why, then, should we be interested in searches in
the near-infrared at all when the optical is so `easy' by comparison (in
terms of the low sky background and the mature detector technology)?
The motivation of moving to the near-infrared is three-fold:
\begin{itemize}
\vspace{-0.25cm}
\item{to fill in the redshift desert at $z\sim1-3$, where most common
emission lines lie outside the optical;}
\vspace{-0.25cm}
\item{to study the rest-frame optical at $z\sim 2$ in order to make a
fair comparison with local Universe -- the rest-optical lines are
vastly more robust to extinction by dust than the rest-UV, with
Ly$\alpha$ particularly unreliable;}
\vspace{-0.25cm}
\item{the near-IR gives us the potential to push to the highest
redshifts yet ($z\ga 10$).}
\vspace{-0.25cm}
\end{itemize}

In this article, I will concentrate on near-infrared search techniques
for actively star-forming galaxies. In their contributions to this
proceedings, Adam Stanford, Peter Eisenhardt and Jim Dunlop discuss
infrared methods of finding and studying high-redshift galaxies with
evolved stellar populations, both in clusters and in the field. Andrew
Blain and Len Cowie describe far-infrared/sub-mm surveys. Throughout I
will consider a cosmology with a vanishing cosmological constant
($\Lambda=0$), $H_{0}=100\,h\,{\rm km\,s}^{-1}\,{\rm Mpc}^{-1}$ and
$q_{0}=0.5$ unless otherwise stated.

\section{Line Emission as an Indicator of Star Formation}

Most high-$z$ emission line searches for field galaxies have focused on
Ly$\alpha$, and until the recent advent of 10-m telescopes have been
uniformly unsuccessful.  With the benefit of hindsight, it appears the
large path-length for the resonant Ly$\alpha$ line in typical neutral
hydrogen columns of $10^{24}\,{\rm m}^{-2}$ greatly enhances the
absorption cross section with only modest dust (\eg, Chen \& Neufeld
1994), depending sensitively on the geometry and kinematics of the
gas. This selective quenching of Ly$\alpha$ is observed in low-$z$
star-bursts (\eg, Kunth \etal\ 1998), and at high redshift Ly$\alpha$
emission is typically weak in the Lyman-limit selected galaxies of
Steidel \etal\ (1996ab,1999) -- indeed this line is actually seen in
absorption in half their sample.  Put simply, the substantial effort in
Ly$\alpha$ searches has yielded mainly upper limits of essentially no
interest for constraining the underlying star-formation rates, as the
extinction of this line is extremely difficult to quantify.

\begin{figure}[h]
%\plottwo{linesnumdens.eps}{sfrlinesz1.eps}
\plottwo{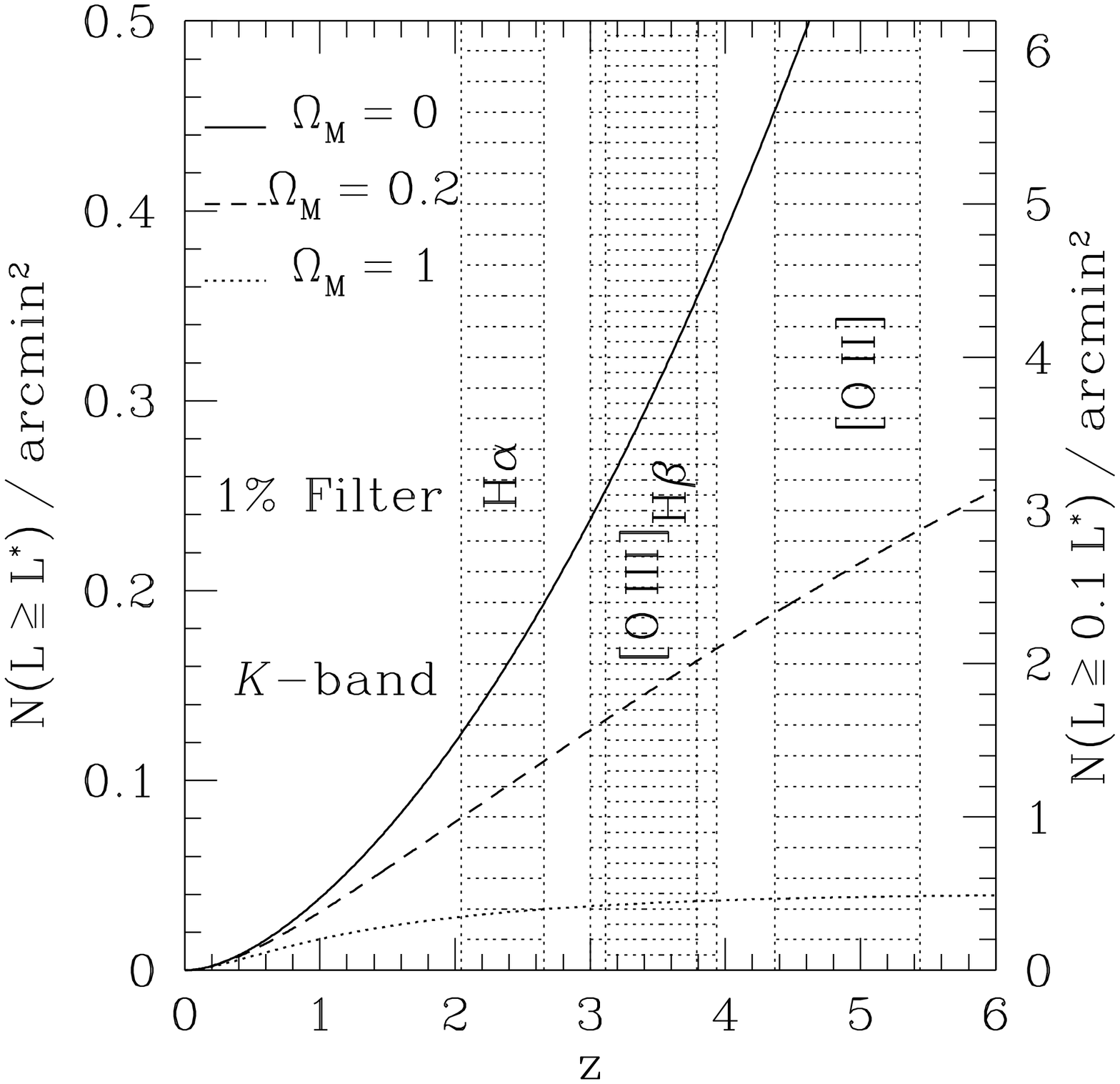}{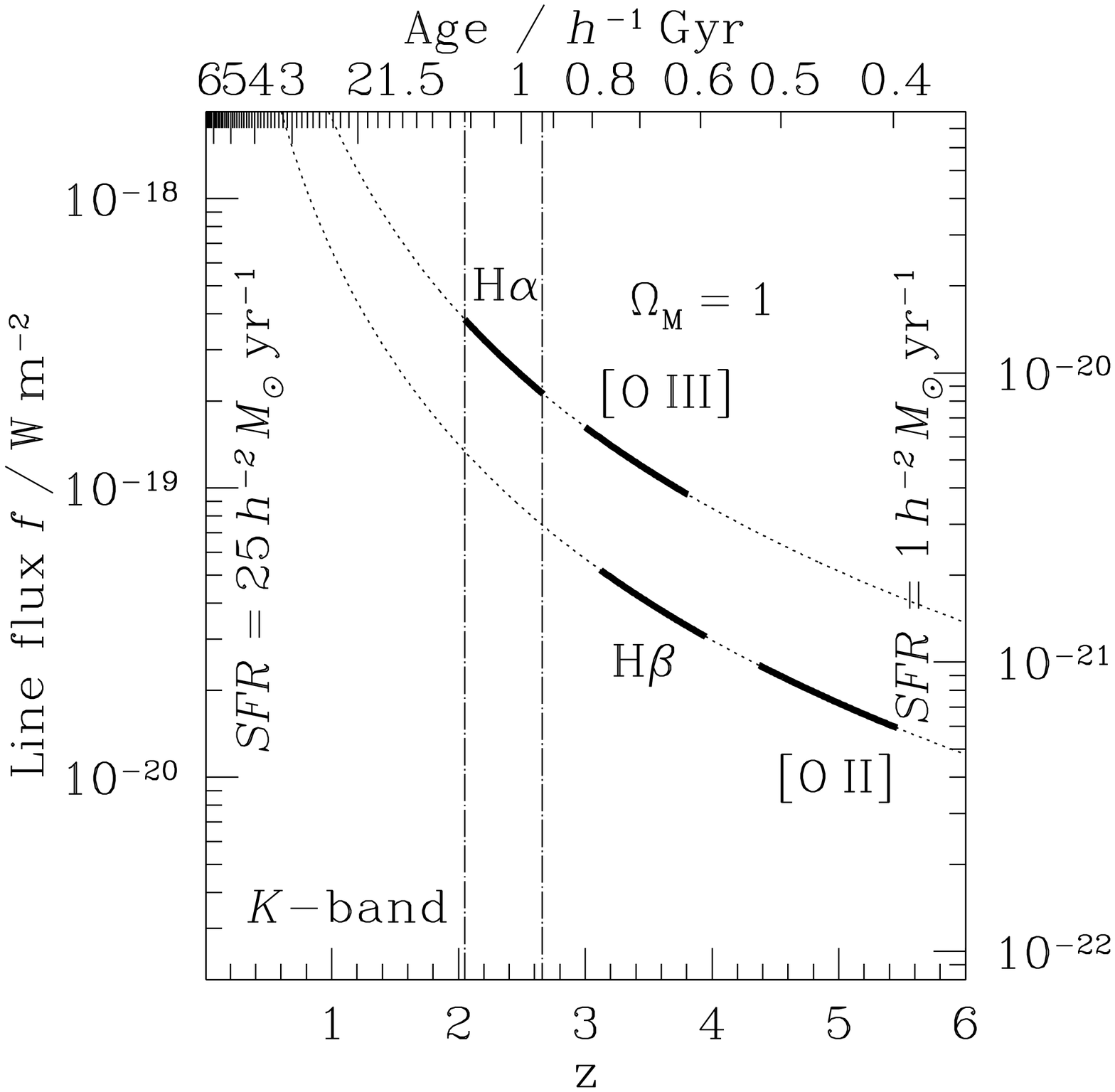}
\caption{{\bf Left:} predictions of the number density of field galaxies
within a 1\% band-pass in the case of no evolution in the co-moving
number density. The surface density brighter than $L^{*}$ (left scale)
or $0.1\,L^{*}$ (right scale) is plotted.  The redshift intervals
covered by four prominent restframe-optical emission lines in the
$K$-band atmospheric window are indicated. Using the $J$- and $H$-bands
too allows complete coverage for $1<z<5$. \newline {\bf Right:} The
predicted (unobscured) line fluxes as a function of redshift for several
emission lines indicative of star formation. The thick
lines indicate the range of redshifts at which these appear in the
$K$-window of the near-infrared. The Kennicutt (1983) relation between
H$\alpha$ luminosity and star formation rate is adopted, and
[O{\scriptsize~II}] is assumed to have a comparable intensity to
H$\beta$ (Kennicutt 1992), and [O{\scriptsize~III}] in the most actively
star-forming systems rivals H$\alpha$ (\eg, Mannucci \& Beckwith 1995),
although the scatter is large.  The left-hand axis shows the line fluxes
for a star formation rate of $25\,h^{-2}\,M_{\odot}\,{\rm yr}^{-1}$,
which may be typical of an $L^{*}$-galaxy progenitor, and the right-hand
axis is for $1\,h^{-2}\,M_{\odot}\,{\rm yr}^{-1}$ (from Bunker 1996).}
\label{fig:numdens}
\label{fig:linefluxz}
\end{figure}

A natural progression is to look for lines of longer rest-frame
wavelength, less affected by extinction than the rest-UV and immune to
the effects of resonant scattering.  The Balmer hydrogen recombination
emission lines, H$\alpha$\,$\lambda$\,656.3\,nm \&
H$\beta$\,$\lambda$\,486.1\,nm, and the collisionally de-excited
forbidden lines of oxygen,
[O{\scriptsize~III}]\,$\lambda\lambda$\,495.9,500.7\,nm \&
[O{\scriptsize~II}]\,$\lambda\lambda$\,372.6,372.9\,nm, are fairly good
indicators of the intrinsic rest-UV continuum from the hottest, most
massive and shortest-lived stars ($M\ga 10\,M_{\odot}$, $T_{\rm MS}\la
10$\,Myr, $T_{\rm eff}\ga 30,000$\,K), and hence the near-instantaneous
star formation rate (SFR). The most fruitful indicator of star formation
at modest redshifts has been the [O{\scriptsize~II}] line.  However,
there is a `dead-zone' between $z\approx 1.3$ (beyond which
[O{\scriptsize ~II}] leaves the optical) and $z\approx 2.7$ (below which
Lyman-break photometric selection is ineffective). Yet some theories
currently in vogue predict this to be the crucial epoch of maximal
merger activity and star formation. Using the atmospheric transmission
windows in the near-infrared -- the $J$ ($1.1-1.4\,\mu$m), $H$
($1.5-1.7\,\mu$m) and $K$ ($2.0-2.4\,\mu$m) pass-bands -- gives access
to the full redshift range $1<z<5$ through the four prominent rest-frame
optical nebular emission lines (H$\alpha$, [O{\scriptsize~III}],
H$\beta$ \& [O{\scriptsize~II}] -- see Fig.~\ref{fig:numdens}).  The
conversion of the nebular line luminosities to star formation rates is
also better understood than that for the sub-mm continuum, which is an
indirect measure of UV flux reprocessed as thermal grain emission in the
far-infrared and depends on a number of factors including the
temperature of the dust, changing by as much as a factor of four over
the range $T_{\rm dust}=30-50$\,K (Barger \etal\ 1999).

%\subsection{Charting the Star Formation History of the Universe}
%
%The history of galaxies -- when and how they formed, and how they have
%evolved -- is a topic of enormous current interest. Building on the work
%of Lilly \etal\ (1996), the plot of Madau \etal\ (1996) endeavours to
%depict the history of global star formation in galaxies from today back
%to $z\ga 4$. However, the current `Madau diagram' is assembled
%principally from optical data and relies on different indicators of star
%formation in the various redshift bins, with uncertain absolute
%calibration and relative susceptibility to dust extinction.  
%
%The
%rest-frame optical emission lines are eminently suitable traces of star
%formation over this range, where they appear in the near-infrared.
%
%Using the $J$-, $H$- \& $K$-bands we can chart the history of star
%formation over the interval $1<z<5$ using the rest-optical nebular
%emission lines alone. These lines are also much less susceptible to
%extinction by dust than the rest-UV continuum and Ly$\alpha$ (which is
%also selectively quenched through resonant scattering).  

\section{Search Strategies for Emission Lines}

A search for line 
emission objects is typically complete to a given line flux and/or
equivalent width limit, 
in contrast to a survey based on broad-band imaging, which is complete 
to a given broad-band magnitude.
%A search for line emission objects does not constitute a
%magnitude-limited survey -- indeed, blank-sky narrow-band imaging and
%long-slit spectroscopy are quite unsuited to this, as the sensitivity to
%line emission is much better than to continuum. 
The goal of such emission line
searches is to trace star formation over a range of redshifts.
While faint magnitude-limited redshift surveys can address this to some
extent, the issue is muddied: selecting on broad-band magnitudes is not
the same as selecting on star formation activity. The rest-frame optical
colours only provide very limited information on the star formation
history. Even the rest-frame UV suffers badly and unpredictably from
dust extinction, and below 121.6\,nm it is gradually eroded by
intervening absorbers.  There may also exist a population of
star-forming galaxies with very large equivalent width emission lines
which would be missing from magnitude-limited surveys on account of
their faint continua, despite high star formation rates (\eg, Curt
Manning's poster paper in this volume). The great advantage of emission
line searches over broad-band magnitude-limited surveys is that they
cleanly pick 
out star forming galaxies at a single redshift, rather than having to
disentangle galaxies at all redshifts down to very different absolute
magnitude limits. The SFR is also measured directly from the nebular
emission lines, with smaller dust corrections than any optical/IR
method.

When constructing a survey for high-redshift star-forming galaxies,
there are six main considerations which must be balanced (\eg, Koo
1986; Djorgovski 1992; Pritchet 1994):
\begin{itemize}
\vspace{-0.25cm}
        \item{{\bf Redshift Coverage:} Does the survey range
cover redshifts where star formation activity is thought to be high? Is
the spread in look-back time adequate to test various models?}
\vspace{-0.25cm}
        \item{{\bf Solid Angle:} Is sufficient area on the
sky covered to intercept several galaxies?}
\vspace{-0.25cm}
        \item{{\bf Clustering:} What density enhancement
above the average for the field can be expected by targeting the
search on known objects at high-$z$?}
\vspace{-0.25cm}
        \item{{\bf Discrimination Against Foreground
Objects:} How effectively can the high-redshift sheep be separated
from low-redshift goats? Is there any foreground contaminant population?}
\vspace{-0.25cm}
        \item{{\bf Sensitivity:} Can faint enough line fluxes
be attained to reach cosmologically interesting star formation rates?}
\vspace{-0.25cm}
        \item{{\bf The Scientific Goals:} How cleanly is the
star-forming population at a particular redshift isolated?}
\vspace{-0.25cm}
\end{itemize}
Unlike broad-band imaging and spectroscopy, narrow-band work samples a
very small wavelength range (and so a small dispersion in
redshifts). This has the advantage of reducing the background noise by
cutting the spectral extent of sky the detector is exposed to, but has
the major drawback that little depth in redshift space is probed
(although targetted searches for clusters may offset this disadvantage:
Mauro Giavalisco and Ray Carlberg discuss the evolution of clustering
properties at high-$z$ in their articles in this
proceedings). Untargetted (``blank-sky'') long-slit spectroscopy can
cover a larger redshift range, but over a very restricted solid
angle. Although covering a large volume, and despite recent advances in
photometric redshift estimation from multi-waveband colours, broad-band
imaging does not offer precise redshift information. It is also
relatively insensitive to line emission due to the high background -- as
is also the case with ground-based slitless spectroscopy (but see
\S\,\ref{sec:grism}).

One of the advantages of high-$z$ searches for H$\alpha$ in particular
is the lack of lower-redshift interlopers to mimic this line emission:
should an object with a single emission line be detected in a $K$-band
search, the most conservative interpretation would be H$\alpha$ at
$z\approx 2-2.5$; as there are no comparably-strong emission lines at
rest-wavelengths longward of 656.3\,nm, it is unlikely that the redshift
is actually less than two (and with plausible higher-redshift
degeneracies of H$\beta$ or [O{\scriptsize~III}]\,500.7\,nm at $z\sim
3.5$ or [O{\scriptsize~II}]\,372.7\,nm at $z\sim 5$). This is in stark
contrast to searches for Ly$\alpha$, where there is frequent confusion
between the high-redshift (Ly$\alpha$-line) interpretation, with a
continuum break at the line attributed to the H{\scriptsize~I} forest
absorption, and a low-redshift galaxy with
[O{\scriptsize~II}]\,372.7\,nm emission accompanied by the
4000\,\AA\,+\,Balmer continuum break. Neither Ly$\alpha$ nor
[O{\scriptsize~II}] have strong nearby lines with which to differentiate
the two interpretations through low-dispersion spectroscopy (see Stern
\etal\ 2000 for a detailed discussion).

\section{Near-Infrared Narrow-Band Imaging}
\label{sec:narrowband}

\begin{figure}[h]
%\resizebox{0.49\textwidth}{!}{\includegraphics{phl957krot.eps}}\hfil
%\resizebox{0.49\textwidth}{!}{\includegraphics{phl957nrot.eps}
\resizebox{0.49\textwidth}{!}{\includegraphics{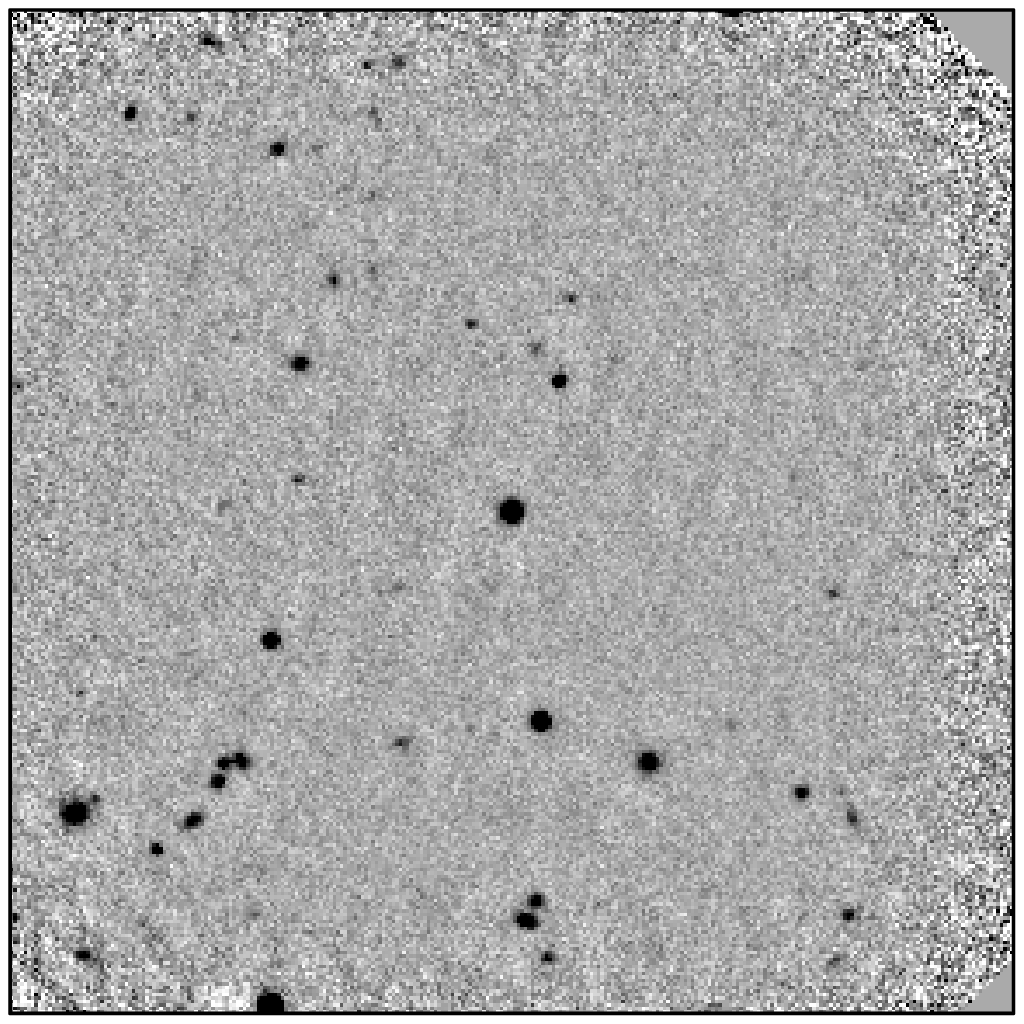}}\hfil
\resizebox{0.49\textwidth}{!}{\includegraphics{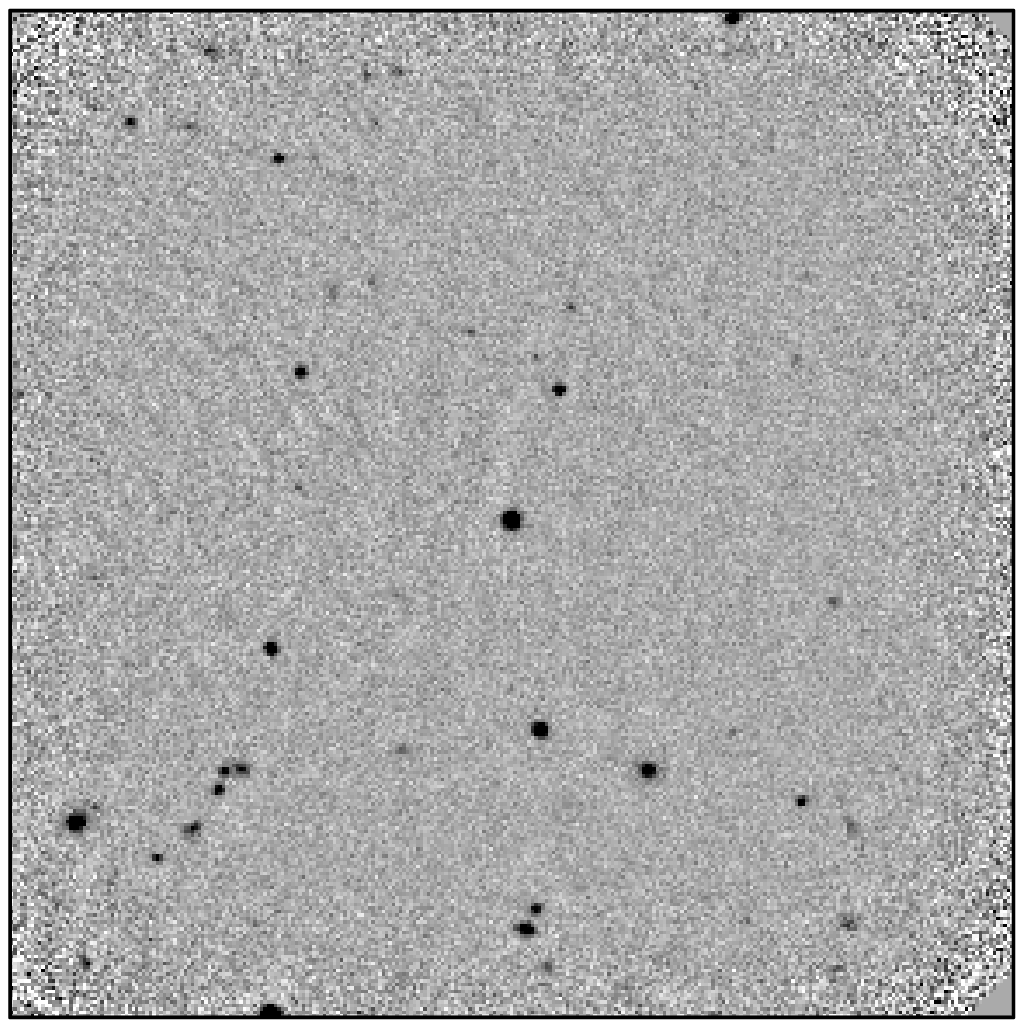}
\put(-436,140){\bf \large $\leftarrow$ PHL\,957}
\put(-359.5,190.2){\circle{20,20}}
\put(-63.5,190.2){\circle{20,20}}
\put(-344,187){\bf \large C1}
\put(-140,140){\bf \large $\leftarrow$ PHL\,957}
\put(-355,260){\huge $K^{\prime}$}
\put(-150,260){\huge $1.75\%\ 2.177\,\mu$m}
\put(-48,187){\bf \large C1}
\put(15,210){\huge \rotatebox{90}{$\longrightarrow$}}
\put(-50,295){\huge $\longleftarrow$}
\put(-450,300){\bf \huge \line(1,0){115}}
\put(-230,300){\bf \huge \line(1,0){48}}
\put(10,245){\bf \huge N}
\put(-80,295){\bf \huge E}
\put(-510,298){\bf \large 1\,arcmin}
\put(-180,298){\bf \large 100\,$h^{-1}$\,kpc}
}
\caption{The field of the quasar PHL\,957, imaged with the $256\times
256$ array IRAC\,2B on the ESO 2.2-m (Bunker \etal\ 1995). The
narrow-band filter (of width $\Delta\lambda\,/\,\lambda=1.75$\%) is tuned to
H$\alpha$ at $z=2.31$, the redshift of the damped Ly$\alpha$ absorber
(Wolfe \etal\ 1986). The broad-band $K^{\prime}$ frame (left) reaches
0.8\,mag deeper than the narrow-band frame (right), but the object C1, a
known galaxy at $z=2.313$ (Lowenthal \etal\ 1991, Hu \etal\ 1993), is
brighter in the narrow-band frame due to the redshifted
H$\alpha+$[N{\scriptsize~II}] line emission.}
\label{fig:phl957image}
\end{figure}

\begin{figure}[h]
%\resizebox{0.49\textwidth}{!}{\includegraphics{mnrasplot.ps}}\hfil
%\resizebox{0.49\textwidth}{!}{\includegraphics{figA2.eps}}
\resizebox{0.49\textwidth}{!}{\includegraphics{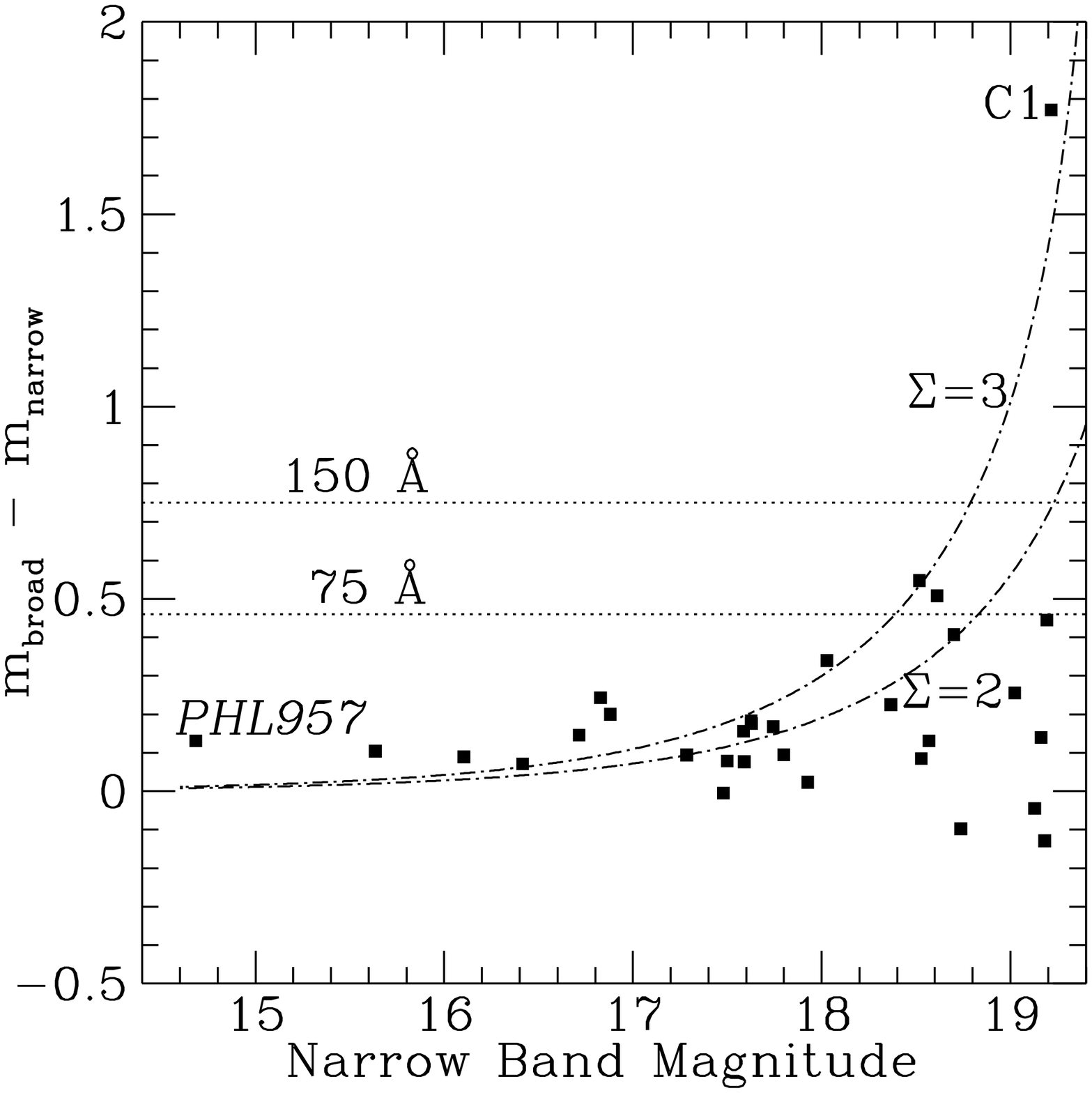}}\hfil
\resizebox{0.49\textwidth}{!}{\includegraphics{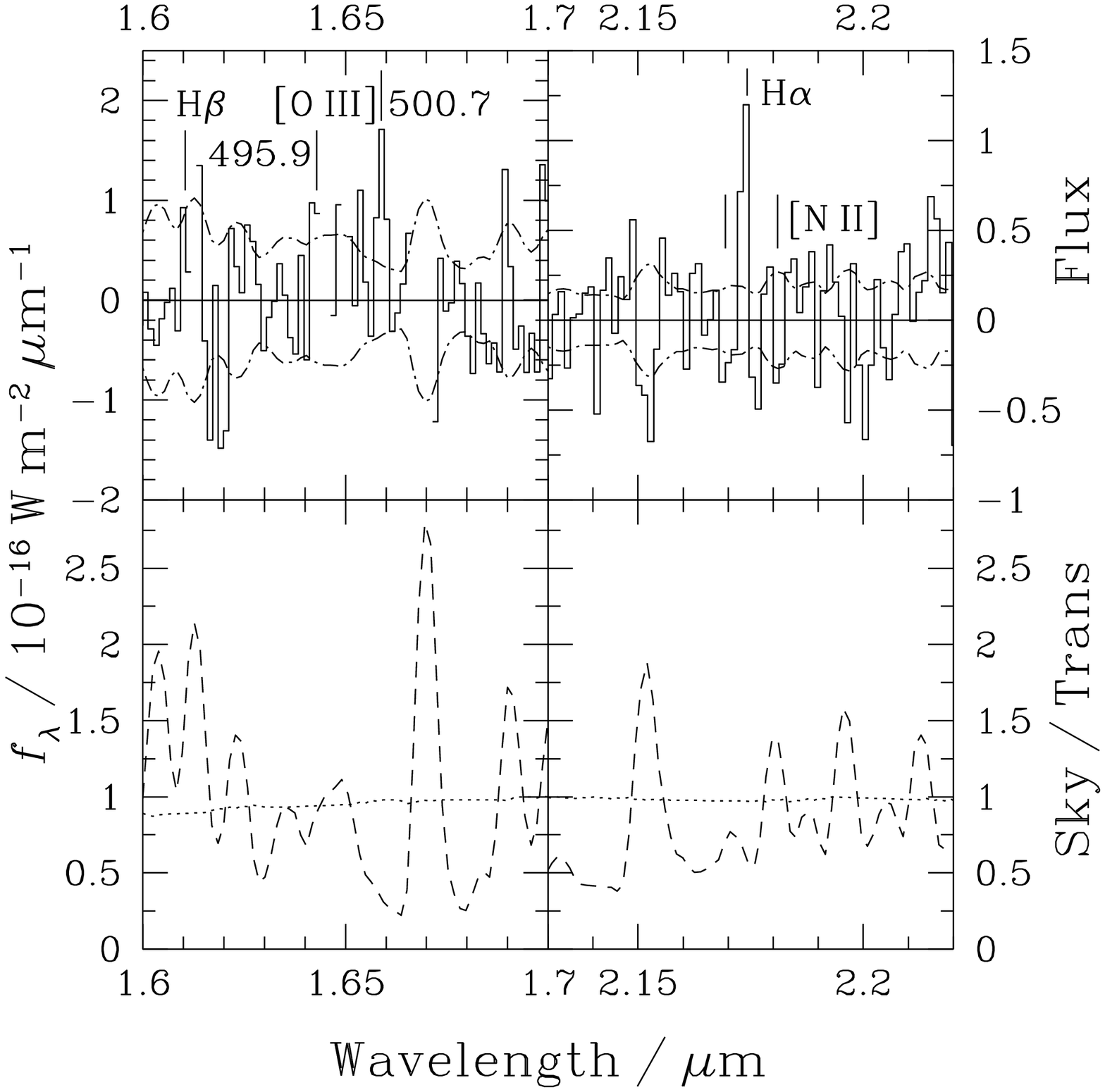}}
\caption{{\bf Left:} Colour-magnitude diagram for the 30 objects
detected at $S/N>4$ in the narrow-band image of the PHL\,957 field by
Bunker \etal\ (1995). The chain-dotted lines are lines of constant
$\Sigma$, which is the number of standard deviations of the excess flux
in the narrow band relative to the broad band.  Also shown are lines of
constant rest-frame equivalent width, for H$\alpha$ at the redshift of
the DLA ($z=2.313$).  Candidate high-redshift galaxies are objects with
equivalent widths ${\rm EW_{rf}>75\,\AA }$, and $\Sigma>2$. The line
$\Sigma=2$ corresponds to a star formation rate of
$11\,h^{-2}\,M_\odot\,{\rm yr}^{-1}$. Note the extreme colour of the
companion, C1 (Fig.~2), due to its H$\alpha$ emission ($f\approx 2\times
10^{-19}\,{\rm W\,m}^{-2}$).
\newline
{\bf Right:} The H$\alpha$ (top-right panel) and
[O{\scriptsize~III}]\,$\lambda$\,500.7\,nm emission (top-left panel)
from C1 (UKIRT/CGS\,4 spectroscopy from Bunker \etal\ 1999). The
dot-dash lines are the $\pm 1\sigma$ noise per pixel, and the lower
panels show the fractional atmospheric transmission (dotted line) and
the sky spectrum scaled down by a factor of 1000 (short-dash line).}
\label{fig:phl957colmag}
\end{figure}

Infrared searches using narrow-band filters (typically ``1\% filters''
with a velocity width of $\approx 3000\,{\rm km\,s}^{-1}$) have become
popular in the last seven years. The technique is to search for objects
which have excess flux in the narrow-band filter when compared to an
off-band, which could be attributable to an emission line being
redshifted into the bandpass (see
Figs.~\ref{fig:phl957image}\,\&\,\ref{fig:phl957colmag}). In practice, a
broad-band filter is commonly used as the `off-band', as the
sensitivity to the continuum is much greater than with simply using an
adjacent narrow-band filter (Pat Hall's paper in this volume describes a
QSO companion detected by this technique).

\subsection{Pilot Studies on 4\,m-Class Telescopes}

One of the first surveys was undertaken on the 3.8-m UK Infrared
Telescope (UKIRT) by Parkes, Collins \& Joseph (1994). This involved
$J$-band imaging with IRCAM-1 (a $62\times 58$ pixel InSb array),
intended to search for Ly$\alpha$ at very high redshift ($7\la z\la 9$)
but also potentially sensitive to H$\alpha$ at $z\approx 0.5-0.9$
(Collins, Parkes \& Joseph 1996).  However, there were no confirmed
line-emission candidates within the $3\,{\rm arcmin}^{2}$ area surveyed
to a limiting flux\footnote{Throughout, I convert the limits quoted for
various surveys to a $3\,\sigma$ threshold in a 3\arcsec -diameter
aperture.} of $10^{-18}-10^{-19}\,{\rm W\,m}^{-2}$. Reaching a
comparable flux limit in the $K$-band, Thompson, Djorgovski \& Beckwith
(1994) surveyed $0.7\,{\rm arcmin}^{2}$ with a similar array on the
Palomar 5-m. This search was targetted on [O{\scriptsize
II}]\,$\lambda$\,372.7\,nm from objects clustered around three $z\ga 4$
QSOs, but did not detect any line-emission companion galaxies.


These pilot studies demonstrated that potentially significant
cosmological volumes and star formation rates could be probed at
high-$z$ through near-infrared narrow-band imaging. However, it was also
clear that deeper surveys covering larger areas were required to unveil
any population of $z>2$ star-forming galaxies.

What should be the minimum survey volume to stand a realistic chance of
intercepting several field galaxies at high-$z$? 
At a redshift of $z\approx 2.3$, where H$\alpha$ appears in the middle
of the $K$-window, a narrow-band survey with a filter of width
$\Delta\lambda\,/\,\lambda$ and a detector of area $\Omega$ will cover a
co-moving volume of:
\[
   \Delta V_{\rm CM} \simeq 10.2\:
\frac{\Delta\lambda\,/\,\lambda}{0.01} \: \frac{\Omega}{1\,{\rm
arcmin}^{2}} \: h^{-3}\,{\rm Mpc}^{3} \: \: (z\approx 2.3,\ q_{0}=0.5).
\]
Locally, the density of field galaxies is $\phi^{*}=0.015\,h^{3}\,{\rm
Mpc}^{-3}$ (Loveday \etal\ 1992)\footnote{This is the same number
density as for the $z\approx 3-4$ Lyman break galaxies of Steidel \etal
(1996ab,1999), although this is probably coincidence: there is unlikely
to be a one-to-one correspondence between today's $L^{*}$ galaxies and
their presumed Lyman break progenitors, as merging will play a major
r\^{o}le.}.  In the idealized case of no evolution in the co-moving
number density, there should be on average one galaxy brighter than
$\,0.3\,L^{*}$ in a volume of $1/\phi^{*}$.  In order to have a 95\%
chance of intercepting {\em at least} one such galaxy, a volume three
times as large should be surveyed (\ie, $200\,h^{-3}$\,Mpc$^{3}$). This
corresponds to 20\,arcmin$^{2}$ with a 1\% filter for $q_{0}=0.5$.  The
number of field galaxies brighter than $1\,L^{*}$ and $0.1\,L^{*}$ per
square arcminute are plotted in Fig.~\ref{fig:numdens} as a function of
redshift.  Any clustering or inclusion of volumes surveyed through
higher-redshift (shorter rest-wavelength) lines serve only to increase
the predicted numbers.

What star formation rates/line fluxes should we expect?  Deep redshift
surveys (\eg, Lilly \etal\ 1996) suggest that $L^{*}$ galaxies have
been mostly assembled by $z\approx 1$, with evolution since then mainly
associated with lower-mass systems (the `down-sizing' of Cowie \etal\
1996).  We concern ourselves here with the quest for the high-redshift
progenitors of the present-day $\approx L^{*}$ galaxies, in the process
of forming a large fraction of their stars: these used to be loosely
referred to as `prim\ae val galaxies' (or `proto-galaxies') before it
was recognised that the star formation history of galaxies was actually
a rather extended and non-coeval process.

At the current epoch, most of the luminous baryonic matter in the
Universe is contained in galaxies with luminosities around $L^{*}$. If
$L^{*}=10^{10}\,h^{-2}\,L_{\odot}$ and a typical stellar mass-to-light
ratio is $M/L = 5\,M_{\odot}/L_{\odot}$ (Faber \& Gallagher 1979) then
to manufacture the mass in stars of an $L^{*}$ galaxy by the current
epoch would require an average star formation rate of $\approx
25\,h^{-2}\,M_{\odot}\,{\rm yr}^{-1}$ over $2$\,Gyr (the Hubble time at
$z\approx 1$ for $q_{0}=0.5$). This is an unobscured line flux of
$f({\rm H\alpha})\sim 3\times10^{-19}\,{\rm W\,m^{-2}}$ at $z\approx
2.3$ (Fig.~\ref{fig:linefluxz}).

This model is, of course, rather simplistic -- both $\phi^{*}$ and
$L^{*}$ will evolve. If merging is important, then we might expect the
high-$z$ progenitors of $L^{*}$ galaxies to be in many pieces. This
hierarchical scenario would increase the surface density of galaxies,
but reduce the average star formation rate per sub-unit -- which would
favour depth rather than area as the pivotal survey
consideration. However, if star formation at high-$z$ is episodic rather
than continuous, then only a fraction of the population may be
detectable in line emission at any given epoch. This has the effect of
reducing the number density of actively star-forming systems, but
because the star formation is concentrated in short bursts interspersed
with quiescent phases, the star formation during a burst must be much
greater than the average value calculated in the simplistic model. If
the star formation history of a galaxy is indeed episodic, this would
drive the survey considerations in the other direction -- towards a
larger area/volume, to maximize the chances of intercepting the luminous
(but infrequent) star-bursts.

\subsection{The First Searches with $256^{2}$ Near-IR Arrays}

The early 1990s saw the emergence of $256^{2}$ InSb and HgCdTe
 near-infrared arrays with low readout noise and background-limited
 performance even with narrow-band filters.  With these, several groups
 undertook surveys which, for the first time, attained the flux limits
 and/or volumes required to test viable models of galaxy formation.  An
 early 10-m Keck program with NIRC placed strong constraints (Pahre \&
 Djorgovski 1995), reaching $2\times10^{-19}\,{\rm W\,m^{-2}}$ over
 4\,arcmin$^{2}$, but did not yield $z>2$ emission-line candidates.  The
 first narrow-band imaging detections of line emission from high-$z$
 objects came soon after.  Bunker \etal\ (1995) imaged in H$\alpha$
 emission a companion of the $z=2.31$ damped Ly-$\alpha$ (DLA) QSO
 absorption system towards the quasar PHL\,957
 (Figs.~\ref{fig:phl957image}\,\&\,\ref{fig:phl957colmag}). Malkan,
 Teplitz \& McLean (1995) also detected H$\alpha$ emission from the
 companion of another QSO absorption system at $z\approx 2.5$ (using
 NIRC/Keck) and subsequently found a cluster (Malkan, Teplitz \& McLean
 1996). In both these cases, the line emission is unlikely to be solely
 due to star formation: the presence of strong high-ionization lines
 such as C{\scriptsize~IV}\,154.9\,nm suggests an AGN contribution.

\begin{figure}[h]
%\plottwo{surlimall.eps}{otherlineslum.eps}
\plottwo{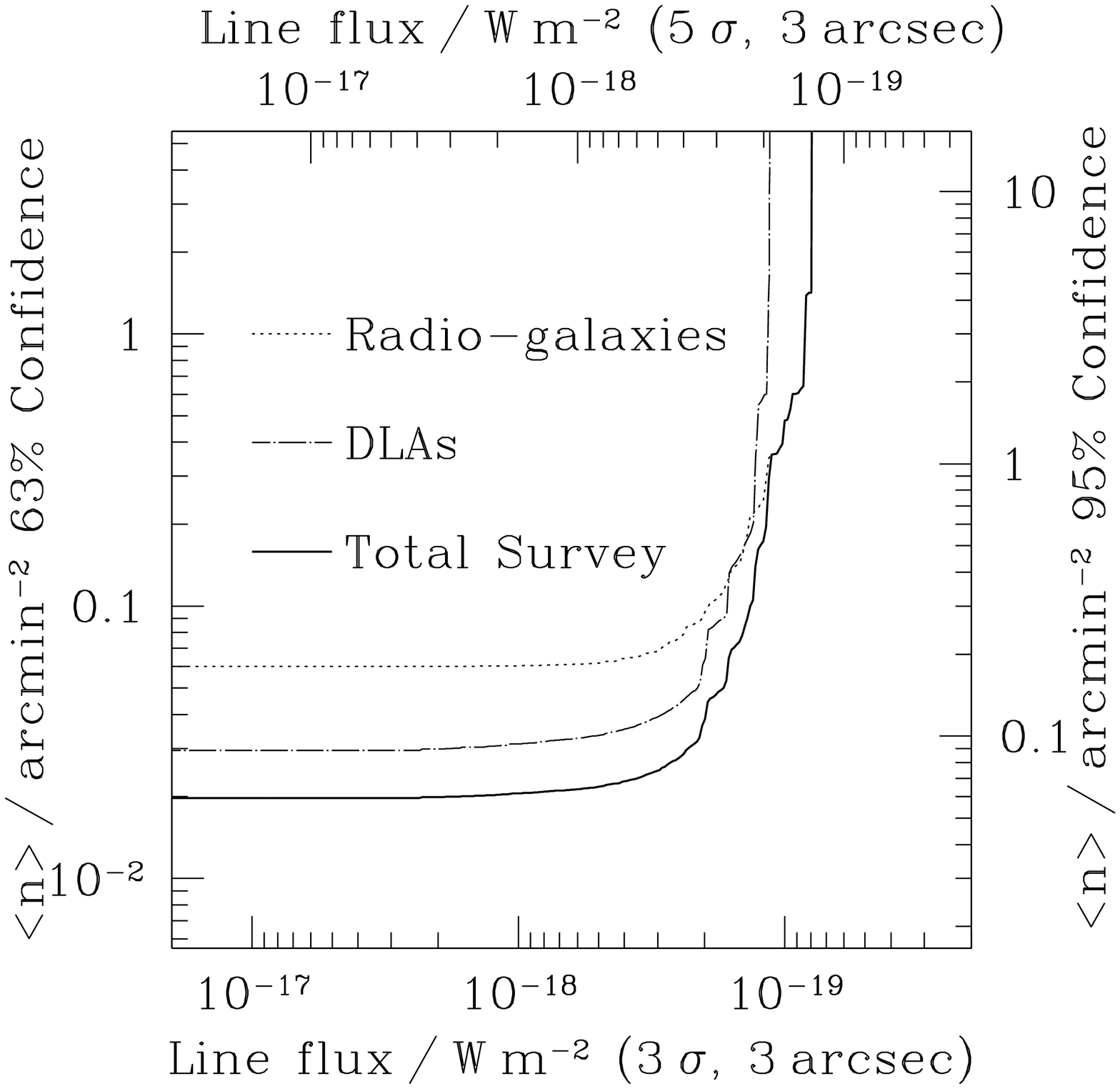}{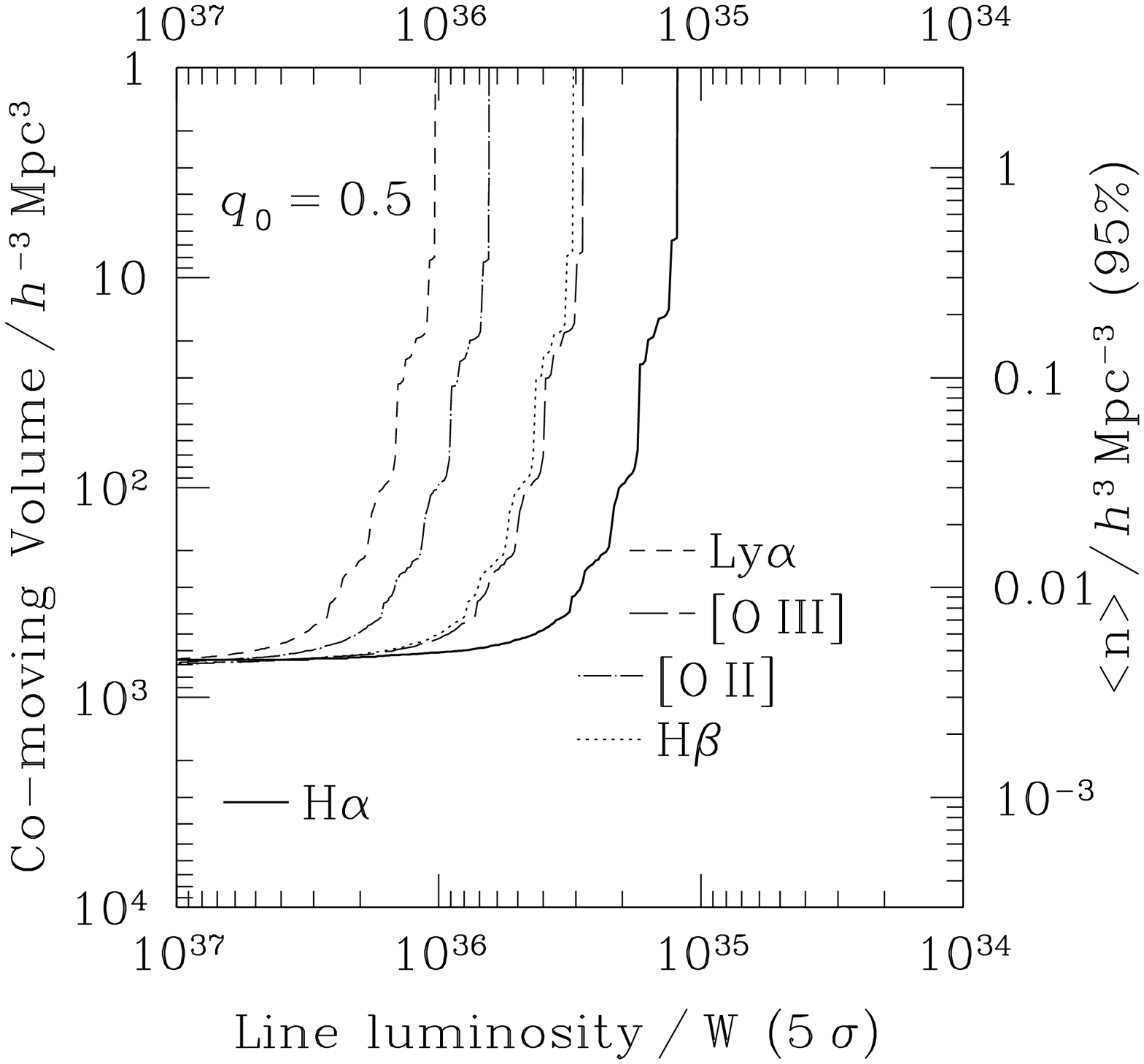}
%\vspace{-0.4cm}
\caption{{\bf Left:} the limits on emission line galaxies placed by the
near-IR narrow-band survey of Bunker (1996), in terms of the observed
quantities: the limiting flux is plotted as a function of the surface
density to which this search is sensitive. Models occupying the region
above and to the left of the line denoting the survey limit are excluded
by the null results of the search. Both the $3\,\sigma$ (lower axis) and
$5\,\sigma$ (upper axis) thresholds in a 3\arcsec -aperture are shown.
Surface densities ruled out at the 63\% (left axis) and 95\% (right
axis) confidence levels are shown. The survey is divided into the total
areas imaged around radio galaxies and damped Ly$\alpha$ systems.
\newline
{\bf Right:} the cumulative co-moving volumes sampled (left axis) and
   $5\,\sigma$ limits on the line luminosities (bottom axis, assuming no
   extinction and $q_{0}=0.5$). The right axis plots the number density
   which would yield on average 3 galaxies in the survey volume: such
   models are excluded at 95\% confidence if they lie above and to the
   left of the line denoting the survey limit for each line (H$\alpha$
   at $z\approx 2.3$; [O{\scriptsize~III}]\,\&\,H$\beta$ at $z\approx
   3.5$, [O{\scriptsize~II}] at $z\sim 5$; and Ly$\alpha$ at $z\sim 16$).}
\label{fig:surlim}
\label{fig:lumlim}
\end{figure}

\begin{figure}
%\plottwo{fig2a.eps}{otherlinesfr.eps}
\plottwo{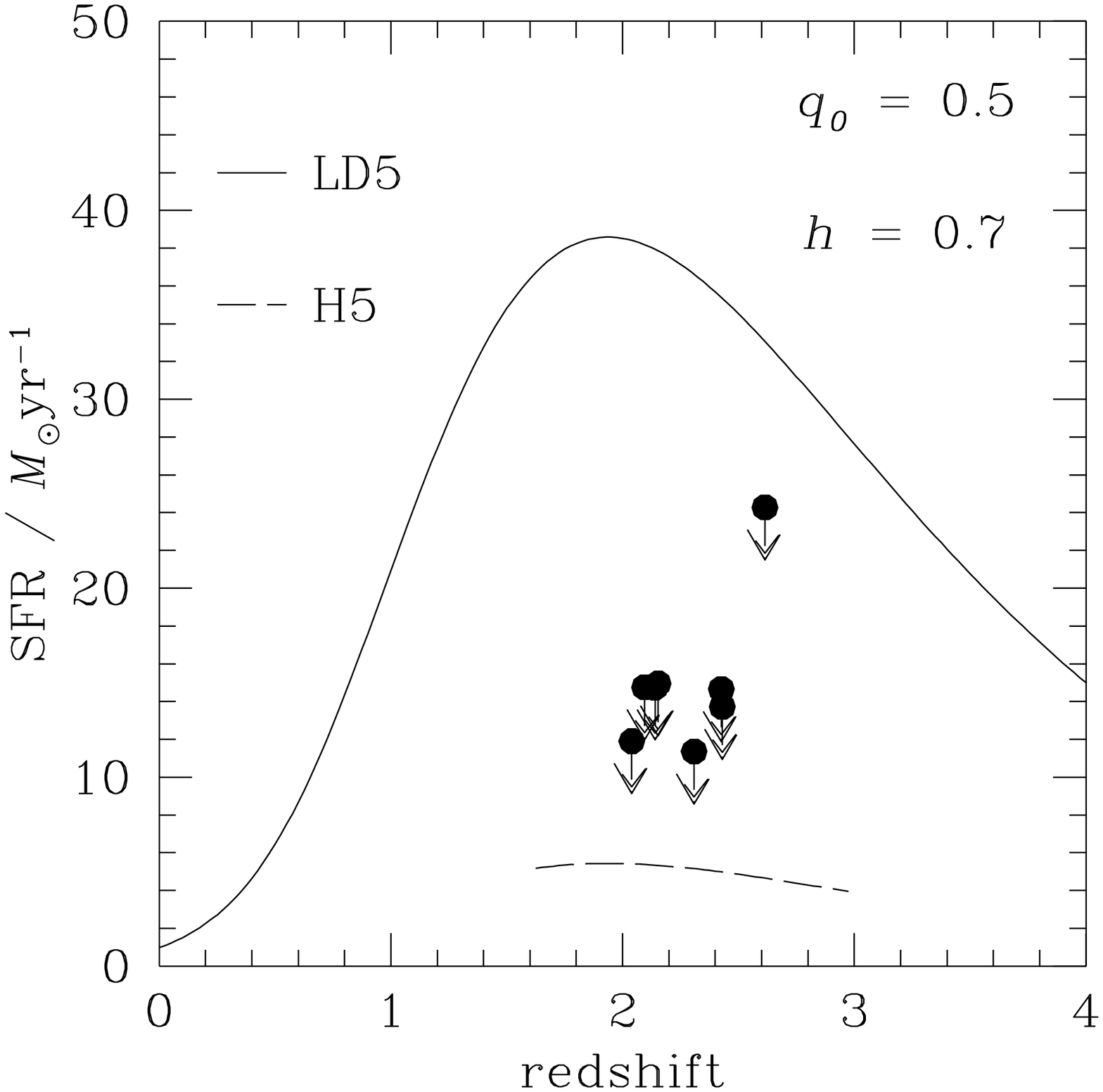}{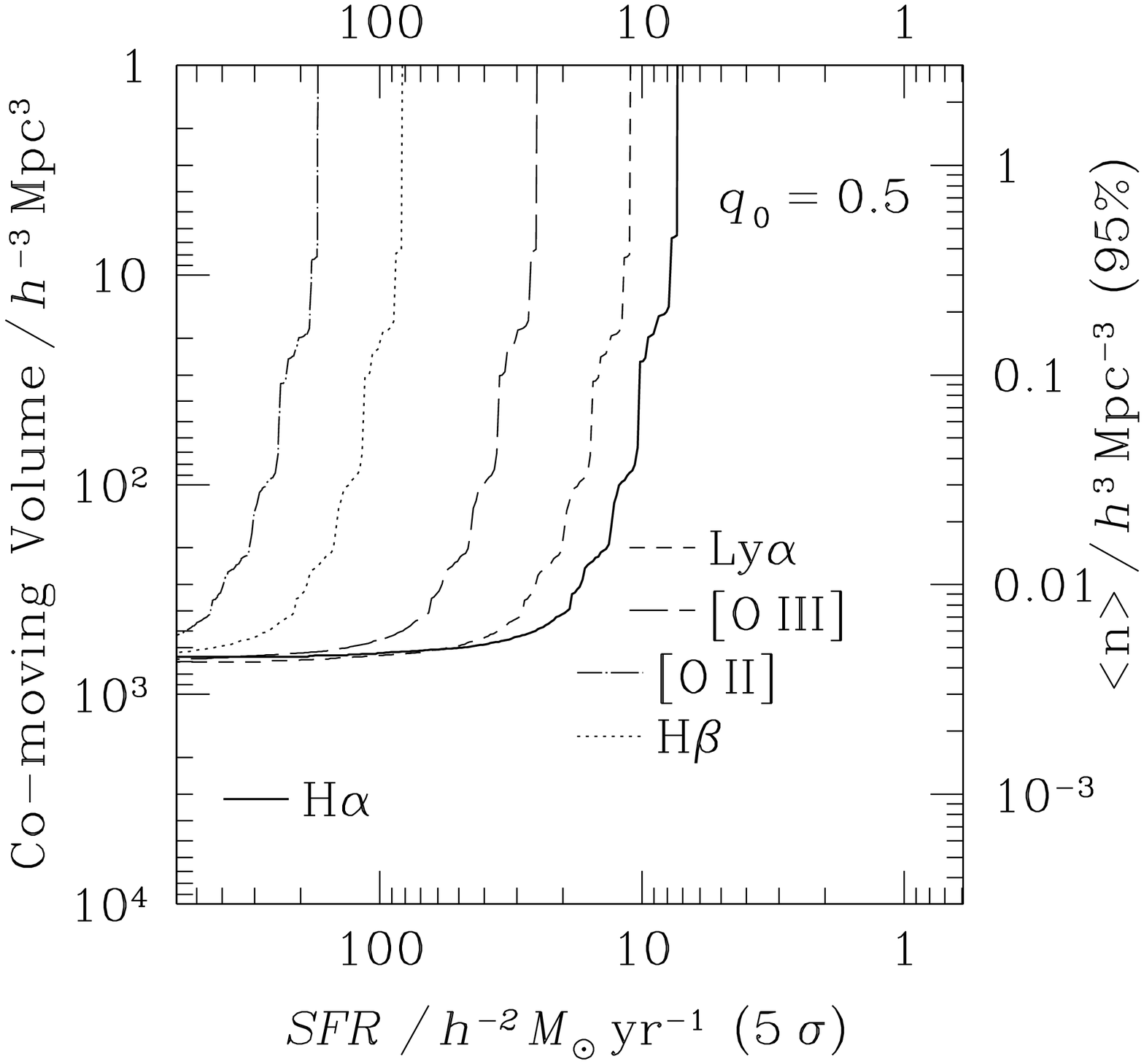}
%\vspace{-0.4cm}
\caption{{\bf Left:} comparison of observational 3$\sigma$ upper limits
(downward arrows) to the SFRs in a sample of 8 DLAs (Bunker \etal\ 1999)
against the predicted sample-averaged (\ie, cross-section weighted) SFRs
for the closed-box models of Pei \& Fall (1995).  The data appear to
rule out the hypothesis that DLAs are the large-disk progenitors of
spiral galaxies in a $q_{0}=0.5$ cosmology (the upper limits fall below
the `LD5' solid curve).  However, the curves plotted take no account of
the possibility that the regions of star formation fall off the
slit. The null detections are still consistent with a hierarchical
picture (the dashed `H5' curve) where the DLAs are sub-galactic units.
\newline {\bf Right:} The upper-limits on the star formation rates
derived from the luminosity limits of various emission lines as a
function of number density (see Fig.~\ref{fig:lumlim}), from the
narrow-band survey of Bunker (1996). Standard case~B unextinguished line ratios
have been adopted -- Ly$\alpha$ is 8 times stronger than H$\alpha$, so
the star formation rates probed are comparable although the redshift is
much higher ($z\approx 16$ {\em vs.\ }$z\approx 2.3$).}
\label{fig:sfrlim}
\label{fig:DLAH5}
\end{figure}

Figs.~\ref{fig:lumlim}\,\&\,\ref{fig:sfrlim} show the constraints on
galaxy evolution from the null results of a typical recent narrow-band
survey (Bunker 1996).  A wider-area search by Thompson, Mannucci \&
Beckwith (1996) used the Calar Alto 3.5-m to survey $\approx 300\,{\rm
arcmin}^{2}$ around QSOs, finding one line-emission source to $3\times
10^{-19}\,{\rm W\,m^{-2}}$ -- an unusual object with broad H$\alpha$ of
$\Delta v_{\rm FWHM}\sim 2000\,{\rm km\,s}^{-1}$, once again probably
powered by an AGN (Beckwith \etal\ 1998). However, there seems to be
large variance: undertaking a comparably-sized survey around
absorption-line systems detected $\sim 20$ candidates (Mannucci \etal\
1998). Using NIRC/Keck to reach fainter limiting fluxes ($\approx
10^{-19}\,{\rm W\,m^{-2}}$) but over a smaller area ($12\,{\rm
arcmin}^{2}$), Teplitz, Malkan \& McLean (1998) report 13 further $z>2$
candidates.

Despite surveying $>500\,{\rm arcmin}^{2}$, narrow-band searches have
not as yet yielded a large population of star forming objects -- there
are $\approx 35$ candidates, of which $\approx 5$ have been
spectroscopically confirmed so far.  Over a similar area ($\sim
1000\,{\rm arcmin}^{2}$), $U$-drop Lyman-break selection by Steidel and
collaborators has yielded 750 spectroscopically-confirmed galaxies. Why
is this?  First, only a thin sliver of redshift-space is sampled within
a narrow-band, compared with that for broad-band colour selection;
second, the infrared searches to date are at the ``tip of the iceberg''
in terms of the luminosity function if dust extinction is modest -- only
the most actively star-forming systems and some AGN are being selected
from current near-IR surveys. For a typical galaxy in Steidel's sample
($M^{*}_{\rm AB}\approx -19.5$ at $\lambda_{\rm rest}=1600$\,\AA ), the
rest-UV would have to be suppressed by a factor of $\approx 7-8$
relative to H$\alpha$ for it to be detected in $K$ at $z\approx 2$
(assuming a typical narrow-band flux limit of $3\times 10^{-19}\,{\rm
W\,m^{-2}}$). This corresponds to a fairly extreme obscuration of
$A_{V}\approx 1.5$\,mag ($A_{V}\la 1$\,mag appears more typical). Given
the small number of galaxies detected so far in H$\alpha$ searches,
there appears not to be a large population of galaxies with
moderately-heavy obscuration. However, narrow-band searches going a
factor of $\approx 2$ deeper over a much wider area are required to
comprehensively test this; the large-format near-infrared arrays on the
new 8-m telescopes make this viable.

\section{Near-Infrared Spectroscopy}
\label{sec:spectra}

\subsection{Slitless Spectroscopy with HST/NICMOS}
\label{sec:grism}

The installation of NICMOS on HST offered an unprecedented opportunity
to chart star formation at $z>1$. The low-background in the $J$- and
$H$-bands compared to ground-based observatories with their bright
atmospheric OH airglow lines, coupled with wavelengths outside the usual
atmospheric windows being accessible, made near-infrared slitless
spectroscopy a truly effective tool for the first time. Comparable flux
limits to the deeper narrow-band searches could be attained in a few
orbits, but the comoving volume surveyed was much greater: the entire
$H$-band was accessible in a single exposure (a range of $z\approx
0.8-1.8$ for H$\alpha$). Pat McCarthy, Lin Yan and colleagues have
analysed the many orbits of parallel-time data obtained with the grisms
on the NIC\,3 array ($52\arcsec\times 52\arcsec$). Surveying several
fields totalling $\approx 65\,{\rm arcmin}^{2}$, McCarthy \etal\ (1999)
found $\approx 30$ single-line emission objects down to a limiting flux
of $0.4\times 10^{-19}\,{\rm W\,m^{-2}}$.  Based on this, Yan \etal\ (1999)
derive an SFR for $z=1.3\pm 0.5$ from H$\alpha$ which is a factor of $3$
higher than that deduced from 280\,nm continua, attributable to dust
extinction (see Lin Yan's contribution to this proceedings for more
details).

\subsection{Near-IR Spectroscopy of Photometrically-Selected Targets}
\label{sec:longslit}

A parallel approach to emission-line searches to determine the global
star formation history is to use photometric redshift estimates from
deep multi-colour imaging, for example the Hubble Deep Fields, to
preselect galaxies likely to be at a suitable redshift (particularly the
`redshift desert' in the optical at $1.3\la z \la 2.7$), and then use
near-infrared spectroscopy to target these and search for the
rest-optical line emission. This technique is potentially very powerful,
as spectroscopy is much more sensitive to line emission than the
narrow-band searches because of the finer resolution element.

With typical resolving powers of $\lambda\,/\,\Delta\lambda_{\rm FWHM} \ga
1000$, the latest near-IR spectrographs can resolve out the OH sky,
enhancing the sensitivity between these lines (particularly in $H$-band,
where the sky continuum is quite dark at wavelengths much shorter than
the thermal infrared).

Glazebrook \etal\ (1999) targetted field galaxies of known redshift from
the CFRS survey (Lilly \etal\ 1996) at $z\approx 1$, and from
UKIRT/CGS\,4 $J$-band spectroscopy inferred a star formation rate from
H$\alpha$ $\approx 3\times$ higher than from the rest UV continuum,
consistent with the NICMOS grism results (\S\,\ref{sec:grism}).

This technique of obtaining near-infrared spectroscopy of galaxies with
previously-established redshifts has recently been extended to the
Lyman-break selected population at $z\approx 3$. Pettini \etal\ (1998)
obtained UKIRT/CGS\,4 spectroscopy with UKIRT on five $U$-drops.  The
recent availability of NIRSPEC on Keck (\eg, James Larkin's paper in
this volume), and the imminent appearance of similar instruments on the
new 8-m telescopes, have the potential to revolutionize this study:
near-infrared spectroscopy of a sample of known $z\sim 3$ galaxies will
shed light on their stellar populations, abundances, true star formation
rates, dust content and kinematics.

The velocity widths measured from the nebular lines are likely to
provide much more reliable kinematic information than Ly$\alpha$, which
is resonantly broadened and typically exhibits a P\,Cygni-like profile
with the blue wing severely absorbed by outflowing neutral hydrogen.
However, the profile of the nebular lines may not be broadened by the
full gravitational potential of the host galaxy; their width may instead
reflect just the velocity dispersion and outflows within the
star-forming H{\scriptsize~II} region (\cf, Pettini \etal\ 1998).

\subsection{Near-Infrared Studies of Damped Absorbers}

An extension of the method of taking a near-infrared spectrum of a known
high-$z$ galaxy to determine its true star formation rate is to study
QSO absorption line systems. We know {\em a priori} that there is a
large gas column density at a particular redshift (causing Ly$\alpha$
absorption of the continuum of the background QSO) which is presumably
associated with a foreground galaxy.  Therefore, a good strategy may be
to target where you think a galaxy is.  In this volume, Varsha Kulkarni
describes a narrow-band search in $H$ with HST/NICMOS for line emission
from a damped Ly$\alpha$ system, and several groups have used
ground-based narrow-band imaging in $K$ to search for H$\alpha$ emission
from DLAs (\eg, Bunker \etal\ 1995; Mannucci \etal\ 1998).

The measurement of quasar absorption lines allows an independent
approach to studying the history of galaxies than the traditional
flux-limited selection. The highest hydrogen column density absorbers
seen in the spectra of background QSOs, the damped Ly$\alpha$ systems
(DLAs), contain most of the neutral gas in the Universe at $z>1$
(Lanzetta \etal\ 1991).  The global history of star formation in the
Universe can be inferred from the evolution in the co-moving density of
neutral gas (derived from the DLA statistics) as it is consumed in star
formation; Pei \& Fall (1995) model this in a self-consistent manner
accounting for dust pollution in the DLAs as star formation
progresses. The average star formation rate in each DLA depends then on
their space density. One school of thought has $z>2$ DLAs being thick
gaseous disks, the progenitors of massive spirals (\eg, Wolfe 1986 and
this proceedings).  Alternatively, DLAs could be more numerous gas-rich
dwarfs, potentially sub-galactic building blocks. To differentiate
between these, my colleagues and I have conducted a search for H$\alpha$
emission from star formation in $z\approx 2.3$ damped systems (Bunker
\etal\ 1999), using near-infrared spectroscopy with CGS\,4 on UKIRT and
building on the previous work of Hu \etal\ (1993).  The absence of any
detectable emission at the faint fluxes probed runs counter to the
predictions of the large disk hypothesis (Fig.~\ref{fig:DLAH5}) --
adding further weight to hierarchical scenarios where today's massive
galaxies were in pieces at high-$z$.

\subsection{The Next Frontier: Lyman-$\alpha$ at $z\ga 10$?}

As we push to even greater redshifts, the optical becomes less and less
useful: the opacity of the intervening H{\scriptsize~I} absorbers
effectively extinguishes most of the flux below $\lambda_{\rm
rest}=121.6$\,nm at $z>5$, forcing a move to the near-infrared.  The
continuum break at Ly$\alpha$ redshift to the near-IR is a potential way
to get to $z>8$, although spectroscopy of the most promising `$J$-drop'
in the HDF-N was inconclusive (Dickinson \etal\ 1999). Despite its poor
track-record, Ly$\alpha$ emission may be a better signature of star
formation in the very early Universe, when chemical enrichment and dust
obscuration were less advanced.  The various $K$-band emission line
searches (\S\,\ref{sec:narrowband} and
Figs.~\ref{fig:lumlim}\,\&\,\ref{fig:sfrlim}) already constrain star
formation at immense redshift ($z\sim 16$); as Avi Loeb's article in
this volume suggests, deep integrations on a 10-m may detect Ly$\alpha$
even before the onset of the Gunn-Peterson effect, with the red-side of
the resonantly-scattered line emission emerging unextinguished.

\section{Conclusions}

There are two primary considerations in the formation and evolution of
galaxies: the assembly of mass (structure formation and the merging
history); and the rate of conversion of neutral gas into stars (the star
formation rate).  Both of these are poorly understood, and may be
regulated by various feedback mechanisms as well as being cloaked by
dust. The current observational constraints are very weak at high
redshift.

 Detection of the rest-optical emission lines in galaxies at $z>2$ is
important to measure the true star formation rates, to correct for dust
and to eliminate systematics in the diagram of evolution in the global
star formation rate (Madau \etal\ 1996; Lilly \etal\ 1996). This
necessitates moving to the near-infrared $J$-, $H$- and $K$-bands.  The
brightness of the infrared sky background and the immature technology
(compared to optical CCDs) has previously been a deterrent to using
these windows. However, the advent of modern detectors with low
read-noise and large format make near-infrared searches for `prim\ae
val' galaxies viable. Near-infrared spectroscopy is about to be
revolutionized through the imminent availability of instruments on
10\,m-class telescopes, and the latest generation of arrays with large
fields-of-view mean that narrow-band searches may at last fulfill their
potential. To push to $z\ga 10$ -- the next frontier in the Hy-redshift
Universe -- demands that we abandon the optical.

\acknowledgments

I am indebted to my main collaborator, Stephen Warren, for his many
insights into the high-redshift Universe.  Steve Rawlings, Mark Lacy,
Gerry Williger, Paul Hewett \& Dave Clements have all been involved in
the near-infrared searches undertaken by the Oxford group. While at
Oxford, my research was supported by a PPARC studentship, and I
received financial support at Berkeley from a NICMOS postdoctoral
fellowship (NASA grant NAG\,5-3043). I gratefully acknowledge
enlightening discussions with Harry Teplitz, Lin Yan, Fillipo Mannucci,
Mike Pahre \& Dave Thompson about their surveys.  Stephen Warren, Mark
Lacy, Francine Marleau, Daniel Stern \& Leonidas Moustakas gave helpful
comments on this manuscript. The hard work of George Djorgovski, Ivan
King \& Daniel Stern made this meeting a reality, and what made it
possible is (of course) the illustrious research career of Hyron Spinrad
-- happy 65th birthday, Hy!

\end{document}